\documentclass[aps,prd,twocolumn,floatfix,nofootinbib,showpacs,superscriptaddress]{revtex4-1}

\usepackage{amsmath,amsfonts,amssymb,bm}
\usepackage{dsfont}
\usepackage{graphicx}
\usepackage{color}
\usepackage{soul}
\usepackage{slashed}

\usepackage[mathscr,scaled=1.15]{urwchancal}
\DeclareFontFamily{OT1}{pzc}{}
\DeclareFontShape{OT1}{pzc}{m}{it}%
{<-> s * [1.15] pzcmi7t}{}
\DeclareMathAlphabet{\mathpzc}{OT1}{pzc}{m}{it}

\definecolor{purple}{rgb}{0.5,0,0.5}
\definecolor{blue}{rgb}{0.0,0,0.9}

\begin{document}

\title{Symmetry preserving truncations of the gap and Bethe-Salpeter equations}

\author{Daniele Binosi}
\affiliation{European Centre for Theoretical Studies in Nuclear Physics
and Related Areas (ECT$^\ast$) and Fondazione Bruno Kessler\\ Villa Tambosi, Strada delle Tabarelle 286, I-38123 Villazzano (TN) Italy}

\author{Lei Chang}
\affiliation{School of Physics, Nankai University, Tianjin 300071, China}

\author{Joannis Papavassiliou}
\affiliation{Department of Theoretical Physics and IFIC, University of Valencia and CSIC, E-46100, Valencia, Spain}

\author{Si-Xue Qin}
\affiliation{Physics Division, Argonne National Laboratory, Argonne, Illinois 60439, USA}

\author{Craig D.~Roberts}
\affiliation{Physics Division, Argonne National Laboratory, Argonne, Illinois 60439, USA}

\date{19 January 2016}

\begin{abstract}
Ward-Green-Takahashi (WGT) identities play a crucial role in hadron physics, \emph{e.g}.\ imposing stringent relationships between the kernels of the one- and two-body problems, which must be preserved in any veracious treatment of mesons as bound-states.  In this connection, one may view the dressed gluon-quark vertex, $\Gamma_\mu^a$, as fundamental.  We use a novel representation of $\Gamma_\mu^a$, in terms of the gluon-quark scattering matrix, to develop a method capable of elucidating the unique quark-antiquark Bethe-Salpeter kernel, $\mathpzc{K}$, that is symmetry-consistent with a given quark gap equation.  A strength of the scheme is its ability to expose and capitalise on graphic symmetries within the kernels.  This is displayed in an analysis that reveals the origin of $H$-diagrams in $\mathpzc{K}$, which are two-particle-irreducible contributions, generated as two-loop diagrams involving the three-gluon vertex, that cannot be absorbed as a dressing of $\Gamma_\mu^a$ in a Bethe-Salpeter kernel nor expressed as a member of the class of crossed-box diagrams.  Thus, there are no general circumstances under which the WGT identities essential for a valid description of mesons can be preserved by a Bethe-Salpeter kernel obtained simply by dressing both gluon-quark vertices in a ladder-like truncation; and, moreover, adding any number of similarly-dressed crossed-box diagrams cannot improve the situation.
\end{abstract}

\pacs{
11.10.St;	
11.30.Rd;	
12.38.Aw;	
12.38.Lg	
}

\maketitle




\noindent\textbf{1.$\;$Introduction}.
A natural framework for studying the two valence-body bound-state problem in quantum field theory is provided by the Dyson-Schwinger equations (DSEs) \cite{Roberts:1994dr}, with the one-body gap equation and two-body Bethe-Salpeter equation (BSE) playing leading roles.  The approach is useful in hadron physics owing to asymptotic freedom in quantum chromodynamics (QCD), which materially reduces model dependence in sound nonperturbative applications because the interaction kernel in each DSE is known for all momenta within the perturbative domain, \emph{i.e}.\ $k^2\gtrsim 2\,$GeV$^2$.  Any model need then only describe the kernels' nonperturbative behaviour.  That is valuable because DSE solutions are propagators and vertices, in terms of which all cross-sections are built.  The approach thus connects observables with the long-range behaviour of QCD's running coupling and masses.  Hence, feedback between predictions and experimental tests can be used to refine any model input and thereby improve understanding of these basic quantities.  This opens the way to addressing questions pertaining to, \emph{e.g}.: the gluon- and quark-structure of hadrons; and the emergence and impact of confinement and dynamical chiral symmetry breaking (DCSB).

The DSEs are a collection of coupled equations; and a tractable problem is only obtained once a truncation scheme is specified.  A weak-coupling expansion reproduces perturbation theory; but, although valuable in the analysis of large momentum transfer phenomena in QCD, it cannot yield nonperturbative information.  A symmetry-preserving scheme applicable to hadrons was introduced in Refs.\,\cite{Munczek:1994zz, Bender:1996bb}.  That procedure generates a BSE from the kernel of any gap equation whose diagrammatic content is known.  It thereby guarantees, \emph{inter} \emph{alia}, that all Ward-Green-Takahashi (WGT) identities \cite{Ward:1950xp, Green:1953te, Takahashi:1957xn, Takahashi:1985yz} are preserved, without fine-tuning, and hence ensures, \emph{e.g}.\ current-conservation and the appearance of Goldstones modes in connection with DCSB.

The leading-order term in the procedure of Refs.\,\cite{Munczek:1994zz, Bender:1996bb} is the rainbow-ladder (RL) truncation.  It is widely used and known to be accurate for light-quark ground-state vector- and isospin-nonzero-pseudoscalar-mesons \cite{Maris:2003vk, Chang:2011vu, Bashir:2012fs, Cloet:2013jya}, and properties of ground-state octet and decuplet baryons \cite{Eichmann:2011ej, Chen:2012qr, Segovia:2013rca, Segovia:2013uga}, because corrections in these channels largely cancel owing to the parameter-free preservation of relevant WGT identities ensured by this scheme.  However, higher-order contributions do not typically cancel in other channels \cite{Roberts:1996jxS, Bender:2002as, Bhagwat:2004hn}.  Hence studies based on the RL truncation, or low-order improvements thereof, usually provide poor results for light-quark scalar- and axial-vector-mesons \cite{Burden:1996nh, Watson:2004kd, Maris:2006ea, Cloet:2007piS, Fischer:2009jm, Krassnigg:2009zh},
exhibit gross sensitivity to model parameters for tensor-mesons \cite{Krassnigg:2010mh} and excited states \cite{Holl:2004fr, Holl:2004un, Qin:2011dd, Qin:2011xq}, and are unrealistic for heavy-light systems \cite{Nguyen:2010yh, Rojas:2014aka, Gomez-Rocha:2015qga}.  

\begin{figure}[t]
\centerline{\includegraphics[width=0.33\textwidth]{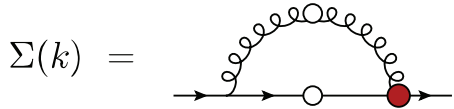}}
\caption{\label{figGap}  Quark self-energy, $\Sigma(k)=i\gamma\cdot k [A(k^2)-1] + B(k^2)$, Eq.\,\eqref{genSigma}: solid line with open circle, dressed-quark propagator $S(q) = 1/[i\gamma\cdot q + \Sigma(q)]$; open-circle ``spring'' , dressed-gluon propagator, $D_{\mu\nu}(p-q)$; and (red) shaded circle, dressed gluon-quark vertex, $\Gamma_\mu^a(q,p)$.  A coupling $g$ appears at each vertex.}
\end{figure}

These difficulties are surmounted by the scheme in Ref.\,\cite{Chang:2009zb} because it enables the use of more realistic kernels for the gap and Bethe-Salpeter equations, which 
possess a sophisticated structure, including Dirac vector$\otimes$vector and scalar$\otimes$scalar quark-antiquark interactions.  Significantly, this technique is also symmetry preserving; but it does not require knowledge of the diagrammatic content of the gap equation's kernel, whose complexity may be expressed in the form chosen for the dressed gluon-quark vertex, a subject of great interest itself, \emph{e.g}.\ Refs.\,\cite{Skullerud:2003qu, Bhagwat:2004kj, Kizilersu:2006et, He:2009sj, Chang:2010hb, Bashir:2011dp, Qin:2013mta, Rojas:2013tza, Aguilar:2014lha}.  The gap equation in Fig.\,\ref{figGap} is:
\begin{equation}
\label{genSigma}
\Sigma(k)= Z_1 \int^\Lambda_{dq}\!\! g^2 D_{\mu\nu}(k-q) \Gamma_\mu^a(k,q)
S(q)\frac{\lambda^a}{2}\gamma_\nu \,,
\end{equation}
where $\int^\Lambda_{dq}$ represents a Poincar\'e invariant regularisation of the four-dimensional integral, with $\Lambda$ the regularization mass-scale, and $Z_{1}(\zeta^2,\Lambda^2)$, is the vertex renormalisation constant, with $\zeta$ the renormalisation scale.
An additional strength of the new scheme is its capacity to express DCSB in the integral equations connected with bound-states.  It has therefore enabled elucidation of novel nonperturbative features of QCD \cite{Chang:2010hb, Chang:2011tx, Chang:2011ei, Chang:2012cc, Chang:2013pq, Chen:2012qr} and facilitated a crucial step toward the \emph{ab initio} prediction of hadron observables in continuum-QCD \cite{Binosi:2014aea}.


Notwithstanding the existence of this improved scheme, there is merit in providing a mechanical approach capable of elucidating that Bethe-Salpeter kernel which is symmetry-consistent with any given gap equation.  Possessing such a tool, one may, \emph{e.g}.\ validate any newly-proposed Bethe-Salpeter kernel that is based on a skeleton expansion of the gap equation, simply by checking whether it ensures preservation of the WGT identities, and/or expose the full complexity demanded of the symmetry-consistent kernel by any concrete statement about the gap equation's structure.  We describe such an approach herein, delivering our explanations mainly in terms of diagrams.  Naturally, each one corresponds to a well-defined integral, which could be written explicitly.  In those terms, however, the proliferation of symbols, nested integrals, etc.\ would obscure the reasoning.  In using diagrammatic methods we are capitalising on the pedagogical capacity and intuitive strengths which have led to Feynman diagrams being adopted so widely.

\smallskip

\begin{figure}[t]
\centerline{\includegraphics[width=0.44\textwidth]{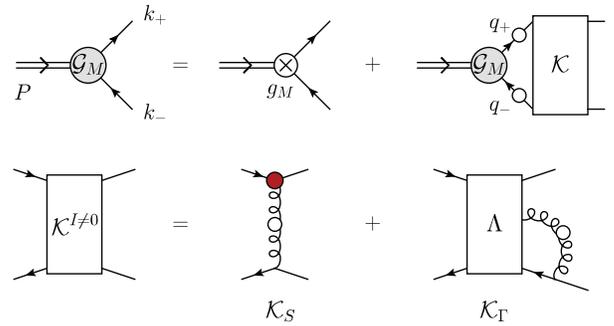}}
%
\caption{\label{FigBSEmeson}  \emph{Upper panel}.  BSE for a colour-singlet vertex, $\mathpzc{G}_M(k;P)$.  The channel is defined by the inhomogeneity, \emph{e.g}.\ with $\mathpzc{g}_M = \tfrac{1}{2} \tau^i \gamma_5\gamma_\mu$ one gains access to all states that communicate with an isovector axial-vector probe, such as the pion and $a_1$ meson.  The interaction between the dressed valence-constituents is completely described by the scattering kernel, $\mathpzc{K}$.
\emph{Lower panel}.   For $I\neq 0$ mesons, the Bethe-Salpeter kernel is a sum of two terms, Eq.\,\eqref{eqKSKG}.  The interaction content of both is completely determined by that of the dressed gluon-quark vertex, explicitly for $\mathpzc{K}_{\;S}$ and implicitly for $\mathpzc{K}_{\;\Gamma}$, Eq.\,\eqref{eqLambda}.
}
\end{figure}

\noindent\textbf{2.$\;$Insufficiency of vertex-dressed ladder kernels}.
The BSE for a colour-singlet vertex, $\mathpzc{G}_M$, which may exhibit meson bound-states, is depicted in Fig.\,\ref{FigBSEmeson}:
\begin{align}
\nonumber
[\mathpzc{G}_M&(k;P)]_{rs}  = Z_M \mathpzc{g}_M + \\
& \int_{dq}^\Lambda [ S(q_+) \mathpzc{G}_M(q;P) S(q_-) ]_{tu} \mathpzc{K}_{tu}^{rs}(k,q;P)\,,
\label{bsetextbook}
\end{align}
where $ Z_M$ is a renormalisation constant,  the total momentum $P=k_+ - k_-$, where $k_+ =  k + \eta P$, $k_- = k - (1-\eta) P$, with $\eta\in[0,1]$: no observable can depend on $\eta$, \emph{i.e}.\ the definition of the relative momentum.

The scattering kernel, $\mathpzc{K}(k,q;P)$ in Eq.\,\eqref{bsetextbook}, expresses all possible interactions that can occur between a dressed quark and dressed antiquark; and is two-particle irreducible (2PI), \emph{viz}.\ it does not contain quark$+$antiquark to single gauge-boson annihilation diagrams nor diagrams that become disconnected by cutting one quark and one antiquark line.  Naturally, this means that $\mathpzc{K}$ also includes an enumerable infinity of $n\geq 2$-PI contributions.

The kernel that ensures preservation of all WGT identities associated with a given colour-singlet vertex may be expressed as \cite{Munczek:1994zz, Bender:1996bb}:
\begin{equation}
\label{EqCalK}
\mathpzc{K}(k,q;P) = -\frac{\delta \Sigma(k)}{\delta S(q)}\,,
\end{equation}
which corresponds to ``cutting'' each internal fermion line in all dressing diagrams.  This cutting procedure actually furnishes a kernel in the ``diagonal configuration'' \cite{Carrington:2012ea}: $\mathpzc{K}(k,q;0)$.  The general momentum configuration may be obtained as described following Eq.\,(44) in Ref.\,\cite{Bender:2002as}, \emph{i.e}.\ in addition to the usual effect of differentiation, the functional derivative adds $P$ to the argument of every quark line through which it is commuted when applying the product rule.  One may alternatively generate the full momentum arguments by beginning with the effective action $\mathpzc{A}[S]$ expressed in coordinate space \cite{Fukuda:1987su, Haymaker:1990vm}, obtaining the dressed-quark propagator as the solution of $\delta{\mathpzc{A}}[S] / \delta S = 0$, and subsequently employing Eq.\,\eqref{EqCalK}.

%

\begin{figure}[t]
\centerline{\includegraphics[width=0.38\textwidth]{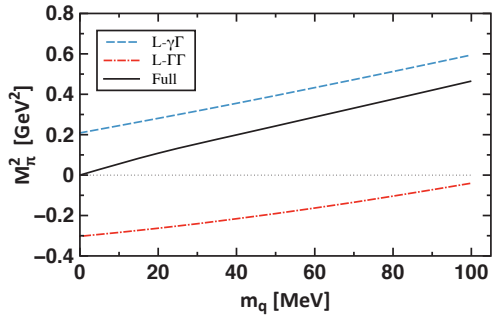}}


%
\caption{\label{Figmpimq}
%
Pion mass-squared vs.\ current-quark mass, obtained with the BC \emph{Ansatz}, Eq.\,\eqref{BCAnsatz}, for $\Gamma_\mu^a$ in the gap equation, Fig.\,\ref{figGap}:
dashed (blue) curve, Bethe-Salpeter kernel in Fig.\,\ref{FigBSEmeson} is $\mathpzc{K}=\mathpzc{K}_{\;S}$;
dot-dashed (red) curve, kernel is $\mathpzc{K}_{\;L\mbox{-}\Gamma\,\Gamma}$, the dressed-vertex ladder-like kernel;
%
and solid (black) curve, complete BC-vertex-consistent kernel constructed following Refs.\,\cite{Chang:2009zb, Qin:2016fbu}.
Evidently, only the complete kernel is sufficient to ensure the existence of a Nambu-Goldstone pion.  (Results obtained with the dressed-gluon line represented by the interaction in Ref.\,\cite{Qin:2011dd}: $D=0.5\,$GeV$^2$, $\omega=0.5\,$GeV, $\tau\to\infty$.)
%
%
}
\end{figure}

For systems with nonzero isospin, $I\neq 0$, quark propagators appearing in gluon vacuum-polarisation diagrams may be neglected and the kernel can be expressed as a sum of just two terms \cite{Bender:1996bb, Bender:2002as, Bhagwat:2004hn, Chang:2009zb, Krein:2014zaa, Heupel:2014ina}:
\begin{equation}
\label{eqKSKG}
\mathpzc{K}^{I\neq 0}(k,q;P) = 
\mathpzc{K}_{\;S}(k,q;P) + \mathpzc{K}_{\;\Gamma}(k,q;P)\,,
\end{equation}
as depicted in the lower panel of Fig.\,\ref{FigBSEmeson}, where the content of the vertex $\Lambda_\mu$ is completely determined by the functional derivative of the dressed gluon-quark vertex:
\begin{equation}
\label{eqLambda}
\Lambda_\mu^a(k,q;P) \sim \frac{\delta \Gamma_\mu^a(k,q)}{\delta S(q)} \,.
\end{equation}
Extending our reasoning to $I=0$ systems is not difficult in principle; but many extra diagrams arise and one must also allow for the possibility that contributions of a topological nature may play an important role \cite{Bhagwat:2007ha}.

\begin{figure}[t]
\centerline{\includegraphics[width=0.47\textwidth]{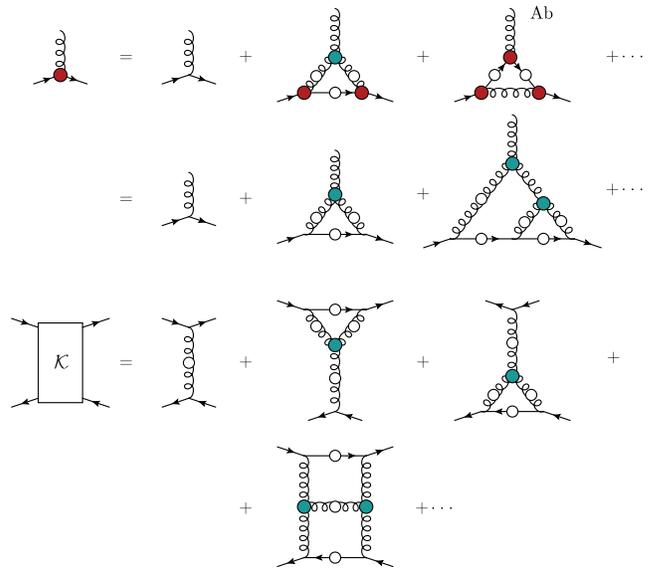}}
%
\caption{\label{FigEqGamma}
\emph{Upper panel}.  DSE for the dressed gluon-quark vertex, $\Gamma_\mu^a$.  Only selected contributions are shown.  The complete equation is depicted, \emph{e.g}.\ in Fig.\,2.6 of Ref.\,\cite{Roberts:1994dr}.  The shaded (blue) circle at the junction of three gluon lines is the dressed three-gluon vertex.
\emph{Lower panel}.  Some of the contributions to the quark-antiquark scattering kernel, $\mathpzc{K}$, generated by the gap equation expressed in terms of the dressed gluon-quark vertex depicted in the upper panel.  The last diagram drawn explicitly is an example of an $H$-diagram.  It is 2PI; but cannot be expressed as a correction to either vertex in a ladder kernel nor as a member of the class of crossed-box diagrams.}
\end{figure}

In some exceptional circumstances, as when $\Gamma_\mu^a$ is expressed via a recursion relation (an integral equation whose kernel has a simple, closed form) and the dressed-gluon propagator has negligible support at nonzero momenta, \emph{e.g}.\ the model in Ref.\,\cite{Munczek:1983dx}, $\Lambda_\mu^a \equiv 0$ in the pseudoscalar channel \cite{Bender:2002as, Bhagwat:2004hn}.  In this case the
Bethe-Salpeter kernel that preserves the axial-vector WGT identity for a given gap equation with dressing defined by $\Gamma_\mu^a$, is obtained by including the specified dressing on only one of the vertices in a ladder-like kernel, \emph{viz}.\ $\mathpzc{K}_{\;S}(k,q;P)$.  That is not true in any other channel and, in general, false in all channels; a fact emphasised by the dashed (blue) curve in Fig.\,\ref{Figmpimq}, which displays the pion mass obtained using a Ball-Chiu (BC) vertex \emph{Ansatz} \cite{Ball:1980ay} $(t=[k+q]/2)$:
\begin{align}
\label{BCAnsatz}
i \Gamma_\mu^a(k,q)  = \frac{\lambda^a}{2} \big[
\varsigma_A i \gamma_\mu  + \delta_A  \tfrac{i}{2} t_\mu \gamma\cdot t
+ \delta_B \, t_\mu \mathbf{I}_{\rm D} \big]\,,
\end{align}
$\varsigma_F = [F(k^2)+F(q^2)]/2$, $\delta_F = [F(k^2)-F(q^2)]/(k^2-q^2)$,
for the dressed vertex in Fig.\,\ref{figGap} and $\mathpzc{K}=\mathpzc{K}_{\;S}$ in the upper panel of Fig.\,\ref{FigBSEmeson}.  Plainly, although the gap equation guarantees DCSB, the BSE does not produce a pion with the characteristics of a Goldstone-boson.

One might imagine that if dressing only one vertex fails, then, perhaps, dressing both vertices, to obtain a dressed-vertex ladder-like kernel, $\mathpzc{K}_{\;L\mbox{-}\Gamma\,\Gamma}$, will be sufficient to produce a symmetry-consistent system.  It is known from Refs.\,\cite{Bender:1996bb, Bender:2002as, Bhagwat:2004hn, Chang:2009zb} that this is false in general: the simplest Abelian-like one-loop gluon-correction to the gluon-quark vertex (generated by the diagram labelled ``Ab'' in Fig.\,\ref{FigEqGamma}) demands the presence of crossed-box contributions to the symmetry-consistent kernel.

%

Suppose though, that one neglects Abelian-like dressings of $\Gamma_\mu^a$; namely, all terms generated by the Ab-diagram in Fig.\,\ref{FigEqGamma} and its analogues in other relevant kernels.  (Abelian-like contributions may be subdominant \cite{Bhagwat:2004hn, Bhagwat:2004kj, Fu:2015tdu}.)
One then arrives at the diagrams in the second line of Fig.\,\ref{FigEqGamma}.  Inserting this vertex into Fig.\,\ref{figGap} and using Eq.\,\eqref{EqCalK}, it is apparent that the first contribution (bare gluon-quark vertex) generates the RL truncation, which is the first term depicted on the right-hand-side (rhs) of the lower panel in Fig.\,\ref{FigEqGamma}.  It is evident, too, that the second diagram in the second line of Fig.\,\ref{FigEqGamma} produces a sum of two terms in $\mathpzc{K}$, \emph{viz}.\ a three-gluon vertex correction on the quark line and another on the antiquark line.  These are the second two terms in the lower panel of Fig.\,\ref{FigEqGamma}; therefore, at this point it might seem plausible that a vertex-dressed ladder kernel can provide a symmetry-preserving framework for the study of mesons.

However, the third diagram drawn in the second line of Fig.\,\ref{FigEqGamma} has not yet been considered.  Amongst others, it generates an $H$-diagram, \emph{viz}.\ the last image drawn in the lower panel of Fig.\,\ref{FigEqGamma}.  Kindred contributions to $\Gamma_\mu^a$ produce an enumerable infinity of similar terms in $\mathpzc{K}$.  Such contributions cannot be expressed as a correction to either vertex in a ladder kernel and neither are they members of the class of crossed-box diagrams.  $H$-diagrams are a distinct class of essentially non-Abelian 2PI contributions to $\mathpzc{K}$.  If they are omitted from the kernel, then the BSE obtained thereby cannot produce colour-singlet vertices that satisfy WGT identities involving a dressed-quark propagator generated by the vertex in Fig.\,\ref{FigEqGamma}, whether or not Ab-type diagrams are neglected.

It is notable that $H$-type diagrams produce an infrared divergence in the perturbative calculation of the static-quark potential, \emph{i.e}.\ a contribution which exhibits unbounded growth as the distance between the source and sink increases \cite{Smirnov:2009fh}.  One might view this as a sign that the inclusion of $H$-type diagrams in the quark-antiquark scattering kernel could be important if one seeks to recover an area-law in the static-quark limit \cite{Brodsky:2015oia}.


The general insufficiency of the vertex-dressed ladder kernel ($\mathpzc{K}_{\;L\mbox{-}\Gamma\Gamma}$) is also highlighted by Fig.\,\ref{Figmpimq}: with the BC vertex, Eq.\,\eqref{BCAnsatz}, one obtains the dot-dashed (red) curve.  Plainly, this kernel generates far too much attraction.  That is not surprising, given the dashed (blue) curve in the same figure, which shows that $\mathpzc{K}_{\;S}$ provides attraction in the pseudoscalar kernel; and, in dressing both vertices, one has not obviously added any repulsion.  Importantly, however, there is actually destructive interference amongst the various contributions in the BSE obtained with $\mathpzc{K}_{\;L\mbox{-}\Gamma\Gamma}$: whereas the $\varsigma_A^2$ terms produce attraction, those involving $\delta_B$ generate net repulsion.

Consider therefore, a modified \emph{Ansatz}, \emph{viz}.\ Eq.\,\eqref{BCAnsatz} with $\delta_B \to 2 \delta_B$.  In this case, weakening the interaction strength: $D=0.5\to 0.29\,$GeV$^2$ produces $m_\pi^2= 0$ at $m_q=0$.  Evidently, one can tune the interaction strength to achieve a massless pion; but securing that numerical outcome is not equivalent to ensuring preservation of the axial-vector WGT identity.  This is readily seen by checking the quark-level Goldberger-Treiman relation \cite{Maris:1997hd, Qin:2014vya}:
\begin{equation}
\label{eqGTRE}
E_\pi(k^2,k\cdot P=0;P^2=0) \stackrel{m_q=0}{\propto} B(k^2)\,,
\end{equation}
a corollary of the axial-vector WGT identity, where $E_\pi$ is the leading term in the pion's Bethe-Salpeter amplitude.  Using the $\delta_B \to 2 \delta_B$ \emph{Ansatz}, both in the gap equation and to generate $\mathpzc{K}_{\;L\mbox{-}\Gamma\Gamma}$, one finds that Eq.\,\eqref{eqGTRE} is violated: an accurate interpolation of the monotonicall decreasing ratio is provided by
$E_\pi(k^2)/B(k^2) = (1+0.02 x)/(1+0.08 x + 0.01 x^2)$, $x\in [0,50]$, $x=k^2/B^2(0)$,
where $B(k^2=0) =0.28\,$GeV.  (We normalised the ratio to unity at $k^2=0$.  It is $0.47$ at $x=9$.)
On the other hand, Eq.\,\eqref{eqGTRE} is preserved without fine tuning when $\mathpzc{K}$ is constructed according to Ref.\,\cite{Chang:2009zb}.  It follows that there exist gap equation kernels, too numerous to count, for which $\mathpzc{K}_{\;L\mbox{-}\Gamma\Gamma}$ yields $m_\pi^2 = 0$ at $m_q=0$; but the pairing nevertheless fails to preserve the axial-vector identity.

The preceding discussion invalidates claims made in Ref.\,\cite{Williams:2015cvx}, \emph{e.g}., referring now to diagrams therein, under no circumstances can the BSE in Fig.\,12 be symmetry-consistent with the gap equation generated by Fig.\,5.




\smallskip

\begin{figure}[t]
\centerline{\includegraphics[width=0.42\textwidth]{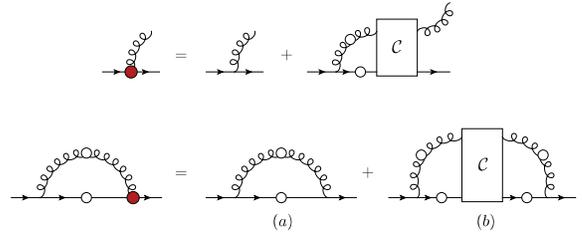}}
%
\caption{\label{fig:VertexSDE}  \emph{Upper panel}.  DSE for the dressed gluon-quark vertex, $\Gamma_\mu^a$, expressed with the antiquark providing the reference line and involving the $s$-channel 1PI gluon-quark scattering amplitude $\mathpzc{C}$.
\emph{Lower panel}.  In terms of the gluon-quark vertex, the equation for the quark self-energy is manifestly symmetric when expressed using $\mathpzc{C}$.}
\end{figure}

\noindent\textbf{3.$\;$Symmetry-consistent Bethe-Salpeter Kernel}.
In order to generalise the discussion in Sec.\,2, we first observe that the common manner of expressing the quark self-energy, Fig.\,\ref{figGap}, is grossly asymmetric with respect to the two gluon-quark vertices: one vertex is fully-dressed, whereas the other has its tree-level form.  This can be remedied by changing the way one looks at the dressed gluon-quark vertex.  Namely, instead of considering the vertex from the gluon's perspective, it is advantageous to adopt the antiquark's view, depicted in Fig.\,\ref{fig:VertexSDE}, and write a DSE for this vertex in terms of the gluon-quark scattering amplitude $\mathpzc{C}$, which is 1PI in the $s$-channel:
\begin{align}
\Gamma_\mu^a(p,q)  & = \tfrac{\lambda^a}{2}\gamma_\mu + \Gamma^{\rm Q}_\mu =: \Gamma^{(0)}_\mu \ast \mathpzc{M}\,, \label{eqM}\\
 \Gamma^{\rm Q}_\mu & =
\int^\Lambda_{d\ell}\!\!
\tfrac{\lambda^b}{2} \gamma_\rho S(\ell_+) D_{\rho\sigma}(\ell_-) \,\mathpzc{C}^{ ba}_{\sigma\mu}(\ell,q;p)\,,
\label{fulvert}
\end{align}
%
%
%
where $\Gamma^{(0)}_\mu=\tfrac{\lambda^a}{2} \gamma_\mu$ is the tree-level contribution, $\Gamma^{\rm Q}_\mu$ expresses all (quantum) corrections, and the sum is represented by the operation of the transition matrix $\mathpzc{M}$, defined implicitly in Eq.\,\eqref{eqM}.  Inserting Eq.\,\eqref{fulvert} into Eq.\,\eqref{genSigma}, one obtains the manifestly symmetric expression for the quark self-energy depicted in the lower panel of Fig.\,\ref{fig:VertexSDE}.

\begin{figure}[t]
\centerline{\includegraphics[width=0.47\textwidth]{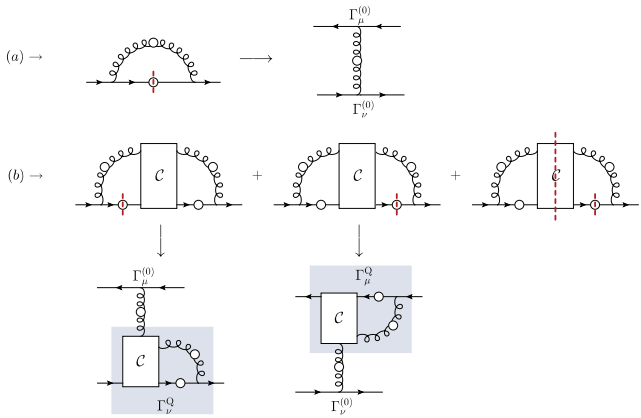}}
\caption{\label{fig:cuts}
In $I\neq 0$ channels, Eq.\,\eqref{EqCalK} produces this series of diagrams for the 2PI kernel of the symmetry-consistent BSE.  The dashed (red) lines represent the act of functional differentiation and the arrows direct attention to the resulting kernel contribution, when that can be depicted simply.}
\end{figure}

Eq.\,\eqref{EqCalK} can now be used to obtain that kernel in Eq.\,\eqref{bsetextbook} which ensures preservation of all WGT identities relevant to the channel considered.  For $I\neq 0$, the differentiation produces the series in Fig.\,\ref{fig:cuts}.  Once more, the rainbow-ladder truncation appears as the simplest contribution, arising from diagram (a) in Fig.\,\ref{fig:VertexSDE}; but it is augmented in general by a series of complex corrections.  Denoting the third diagram in line (b) of Fig.\,\ref{fig:cuts} by $\Sigma_{\not \mathpzc{C}}(k)$, \emph{viz}.\ the image with $\mathpzc{C}$ itself being cut, and represents its contribution in the BSE by $\mathpzc{K}_{\;\not \mathpzc{C}}$, then it is evident from Fig.\,\ref{fig:cuts} that
\begin{align}
\nonumber
\mathpzc{K}^{I\neq 0}  & =- g^2 \big[  \Gamma^{(0)}_\mu \, D_{\mu\nu} \, \Gamma^{(0)}_\nu\\
&
\quad + \Gamma^{(0)}_\mu \, D_{\mu\nu} \, \Gamma^{\rm Q}_\nu
+ \Gamma^{\rm Q}_\mu \, D_{\mu\nu} \, \Gamma^{(0)}_\nu \big]
+\mathpzc{K}_{\;\not \mathpzc{C}}\\
&= \underbrace{- g^2 \Gamma_\mu D_{\mu\nu} \Gamma_\nu}_{\mathpzc{K}_{\;L-\Gamma\,\Gamma}}
+ g^2 \Gamma^{\rm Q}_\mu D_{\mu\nu} \, \Gamma^{\rm Q}_\nu + \mathpzc{K}_{\;\not \mathpzc{C}}\,.
\label{eqKernelOGE}
\end{align}

\begin{figure}[t]
\centerline{\includegraphics[width=0.47\textwidth]{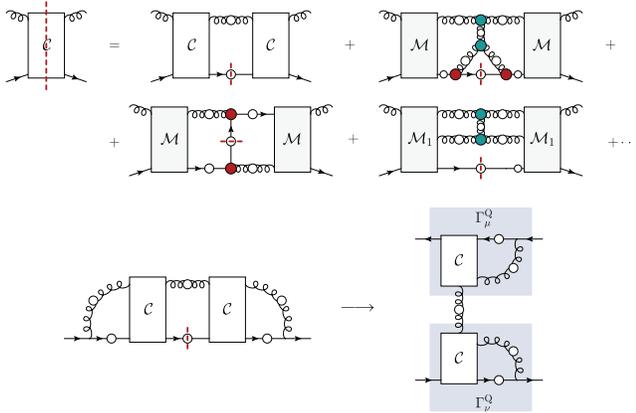}}
%
\caption{\label{figdMpdS}
\emph{Upper panel}.  In $I\neq 0$ channels, $\mathpzc{K}_{\;\not \mathpzc{C}}$ has the expansion depicted here, where the dashed (red) lines indicate the quark line that reacts to the functional differentiation in Eq.\,\eqref{EqCalK}.  The expansion necessarily involves $g+q \to (m+1) g +q $, $m\geq 1$, transition matrices, $\mathpzc{M}_1$, etc.
\emph{Lower panel}.  A focus on the first diagram on the rhs in the upper panel reveals the simple nature of its contribution to $\mathpzc{K}$.}
\end{figure}

Plainly, $\mathpzc{K}_{\;L\mbox{-}\Gamma\Gamma}$ in Eq.\,\eqref{eqKernelOGE} is a dressed-vertex ladder-like subcomponent of the symmetry-consistent kernel (already considered in Sec.\,2); but there is much more in the complete kernel.
To elucidate, we focus on $\mathpzc{K}_{\;\not \mathpzc{C}}$ and consider the nature of $\mathpzc{C}$.  This amplitude contains infinitely many diagrams; and although an infinite number of them contain no internal quark line, that collection cannot contribute to $\mathpzc{K}_{\;\not \mathpzc{C}}$.  The relevant contributions to $\mathpzc{C}$ are those which contain at least one quark line; and hence, for $I\neq 0$, $\mathpzc{K}_{\;\not \mathpzc{C}}$ has the expansion depicted in Fig.\,\ref{figdMpdS}.

\begin{figure}[t]
\centerline{\includegraphics[width=0.25\textwidth]{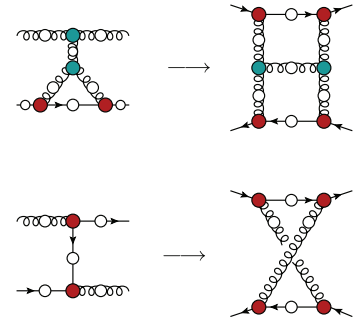}}
\caption{\label{figExplicitH}
Explicating the origin of $H$-diagrams and crossed-box terms in the quark-antiquark scattering kernel, $\mathpzc{K}^{I\neq 1}$.  \emph{Left} -- elements in the gluon-quark scattering matrix, $\mathpzc{C}$; and \emph{right} -- contributions they generate in $\mathpzc{K}_{\;\not \mathpzc{C}}$.  }
\end{figure}

The lower panel of Fig.\,\ref{figdMpdS} focuses on the first term on the rhs of the upper panel, which produces the following contribution to the quark-antiquark kernel:
$(-g^2)\Gamma^{\rm Q}_\mu D_{\mu\nu} \Gamma^{\rm Q}_\nu$.
All remaining terms generate $n\geq 2$-PI contributions to $\mathpzc{K}^{I\neq 0}$ that are structurally inequivalent to those contained in $\mathpzc{K}_{\;L\mbox{-}\Gamma\,\Gamma}$, as illustrated by Fig.\,\ref{figExplicitH}.  Denoting these terms by  $\mathpzc{K}_{\;\not L}$, one arrives finally at
\begin{equation}
\label{eqKLnotL}
\mathpzc{K}^{I\neq 0}  = 
\mathpzc{K}_{\;L\mbox{-}\Gamma\,\Gamma} + \mathpzc{K}_{\; \not L} \,.
\end{equation}
That it is impossible for $\mathpzc{K}_{\;L\mbox{-}\Gamma\,\Gamma}$ alone to serve as a symmetry-consistent quark-antiquark scattering kernel is also evident here.  Amongst infinitely many others, the second and third images drawn explicitly in the expression for $\mathpzc{K}_{\;\not \mathpzc{C}}$ in Fig.\,\ref{figdMpdS} contain the contributions in Fig.\,\ref{figExplicitH}: the top-left image expresses a correction to the gluon-quark vertex, so its influence is felt within $\mathpzc{K}_{\;L\mbox{-}\Gamma\,\Gamma}$; but it must simultaneously contribute to $\mathpzc{K}_{\; \not L}$, as displayed in Fig.\,\ref{figdMpdS}, producing the $H$-diagram depicted Fig.\,\ref{figExplicitH}.

\smallskip

\noindent\textbf{4.$\;$Epilogue}.
Working with a simple \emph{Ansatz} for the dressed gluon-quark vertex, $\Gamma_\mu^a$, we considered the capacity of vertex-dressed ladder-like Bethe-Salpeter kernels to preserve Ward-Green-Takahashi (WGT) identities relevant to meson bound-states; and found that whilst they can readily be tuned to produce a massless pion in the chiral limit, they nevertheless fail to preserve the axial-vector WGT identity and are thus incomplete.
We generalised these observations using a novel representation of $\Gamma_\mu^a$ in terms of the gluon-quark scattering amplitude, which enabled us to show that whilst a dressed-vertex ladder-like truncation is the simplest term in the complete Bethe-Salpeter kernel, $\mathpzc{K}$, it is insufficient in general, owing to the presence of, \emph{inter alia}, $H$-diagrams, \emph{viz}.\ two-loop terms, involving the three-gluon vertex, that cannot be absorbed into $\mathpzc{K}$ as a part of $\Gamma_\mu^a$ nor expressed as a member of the class of crossed-box diagrams.  Consequently, the WGT identities essential for a valid description of mesons cannot generally be preserved when $\mathpzc{K}$ is obtained merely by dressing both gluon-quark vertices in a ladder-like truncation; and, moreover, adding any number of similarly dressed terms in the class of crossed-box diagrams cannot improve the situation.  Fortunately, sophisticated alternatives exist \cite{Chang:2009zb}, are in practical employment \cite{Chang:2010hb, Chang:2011tx, Chang:2011ei, Chang:2012cc, Chang:2013pq, Binosi:2014aea} and are being refined \cite{Qin:2016fbu}.
\smallskip

\noindent\textbf{Acknowledgments}.
We are grateful for constructive remarks from
A.~Bashir,
S.\,J.~Brodsky,
I.\,C.~Clo\"et,
B.~El-Bennich,
G.~Krein,
C.~Mezrag,
J.~Rodr\'{\i}guez-Quintero,
and J.~Segovia.
Research supported by:
Spanish MEYC grant nos.\ FPA2014-53631-C2-1-P and SEV-2014-0398;
Generalitat Valenciana under grant “PrometeoII/2014/066”;
Argonne National Laboratory Office of the Director, through the Named Postdoctoral Fellowship Program;
and U.S.\ Department of Energy, Office of Science, Office of Nuclear Physics, contract no.~DE-AC02-06CH11357.




\end{document}